\documentclass[9pt,conference]{IEEEtran}

\usepackage{amssymb,amsthm,amsmath,array}
\usepackage[mode=buildnew]{standalone}
\usepackage{graphicx}
\graphicspath{{Images/}}
\usepackage{xspace}
\usepackage[sort&compress, numbers]{natbib}
\usepackage{stmaryrd}
\usepackage{xcolor}
\usepackage{mathtools}
\usepackage{float}
\usepackage{textcomp}
\usepackage{caption}
\usepackage{subcaption}
\usepackage{gensymb}
\usepackage{tikz}
\usepackage{hyperref}
\usepackage{pgfplots}

\newcommand{\ie}{\emph{i.e.},\xspace}
\newcommand{\eg}{\emph{e.g.},\xspace}

\newcommand{\sq}{\vspace{-1.5mm}}

\begin{document}
\title{Subsampling Methods for Fast Electron Backscattered Diffraction Analysis }
\author{\IEEEauthorblockN{
        Zoë Broad\IEEEauthorrefmark{1}, 
        Daniel Nicholls\IEEEauthorrefmark{1},
        Jack Wells\IEEEauthorrefmark{2},
        Alex W. Robinson\IEEEauthorrefmark{1},
        Amirafshar Moshtaghpour\IEEEauthorrefmark{1}\IEEEauthorrefmark{3},\\
        Robert Masters\IEEEauthorrefmark{4},
        Louise Hughes\IEEEauthorrefmark{4}, and 
        Nigel D. Browning\IEEEauthorrefmark{1}
    }
    \IEEEauthorblockA{
        \IEEEauthorrefmark{1} Department of Mechanical, Materials and Aerospace Engineering, University of Liverpool, UK.\\
        \IEEEauthorrefmark{2} Distributed Algorithms Centre for Doctoral Training, University of Liverpool, Liverpool, UK.\\
        \IEEEauthorrefmark{3} Correlated Imaging Group, Rosalind Franklin Institute, Harwell Science and Innovation Campus, Didcot, UK.\\
        \IEEEauthorrefmark{4} Oxford Instruments Nanoanalysis, High Wycombe, UK.}
        }
\maketitle
\begin{abstract}
    Despite advancements in electron backscatter diffraction (EBSD) detector speeds, the acquisition rates of 4-Dimensional (4D) EBSD data, \ie a collection of 2-dimensional (2D) diffraction maps for every position of a convergent electron probe on the sample, is limited by the capacity of the detector. 
    Such 4D data enables computation of, \eg band contrast and Inverse Pole Figure (IPF) maps, used for material characterisation. In this work we propose a fast acquisition method of EBSD data through subsampling 2-D probe positions and inpainting. We investigate reconstruction of both band contrast and IPF maps using an unsupervised Bayesian dictionary learning approach, \ie Beta process factor analysis.
    Numerical simulations achieve high quality reconstructed images from 10\% subsampled data.
\end{abstract}

\section{Introduction}
Electron backscatter diffraction (EBSD) is a scanning electron microscopy based technique used for the characterization of microstructures, providing crystallographic information such as grain morphology and crystal orientation \cite{code1}.  

\begin{figure}
    \centering
    \includegraphics[scale=0.4]{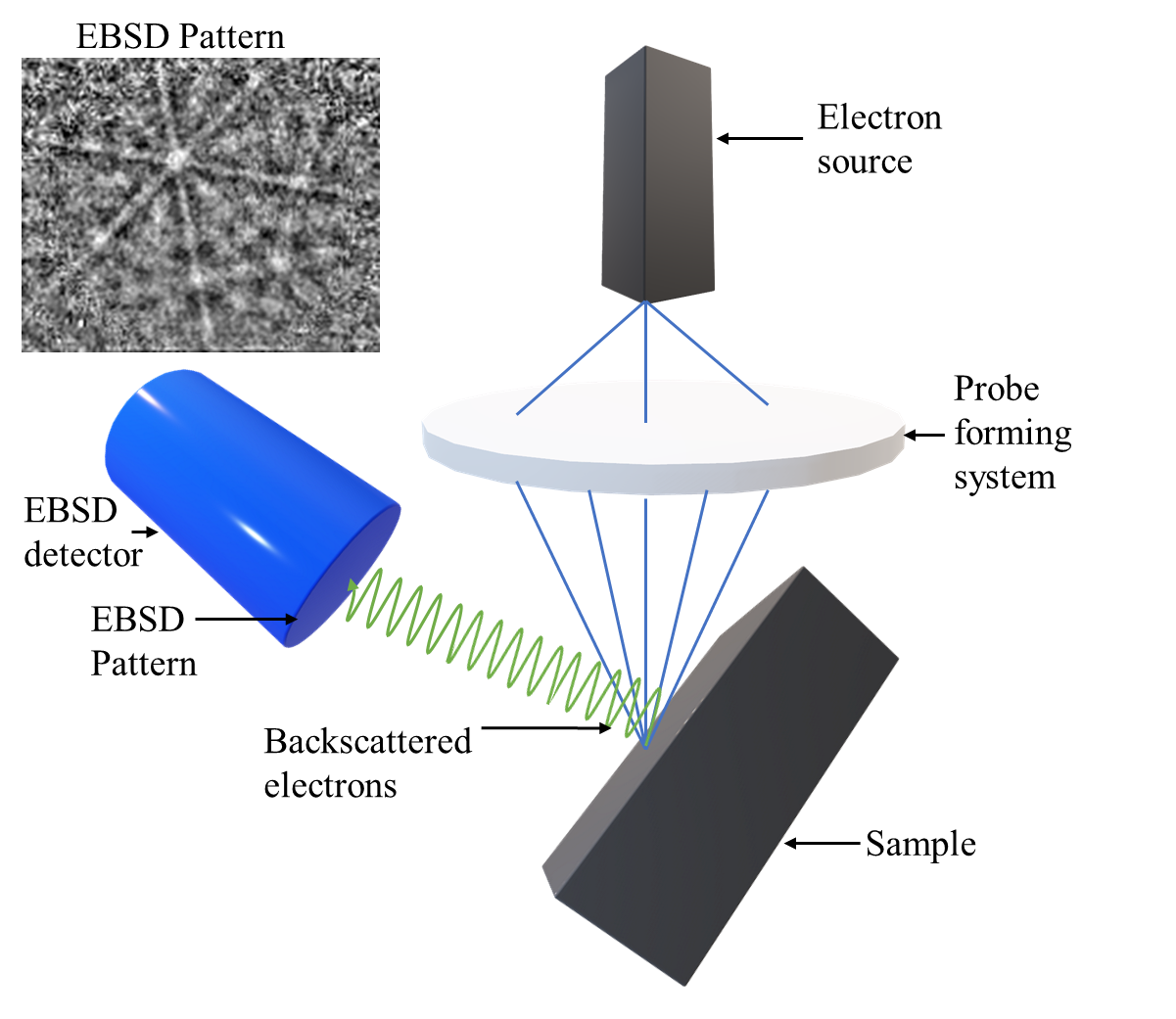}
    \caption{Schematic of EBSD Setup. A convergent electron beam is raster scanned across the sample. Backscattered electrons form a pair of cones which intersect the phosphor screen, allowing the pattern to be read by the detector.}
    \label{fig:SetUp}
\end{figure}
\begin{figure}
    \centering
    \includegraphics[scale=0.6]{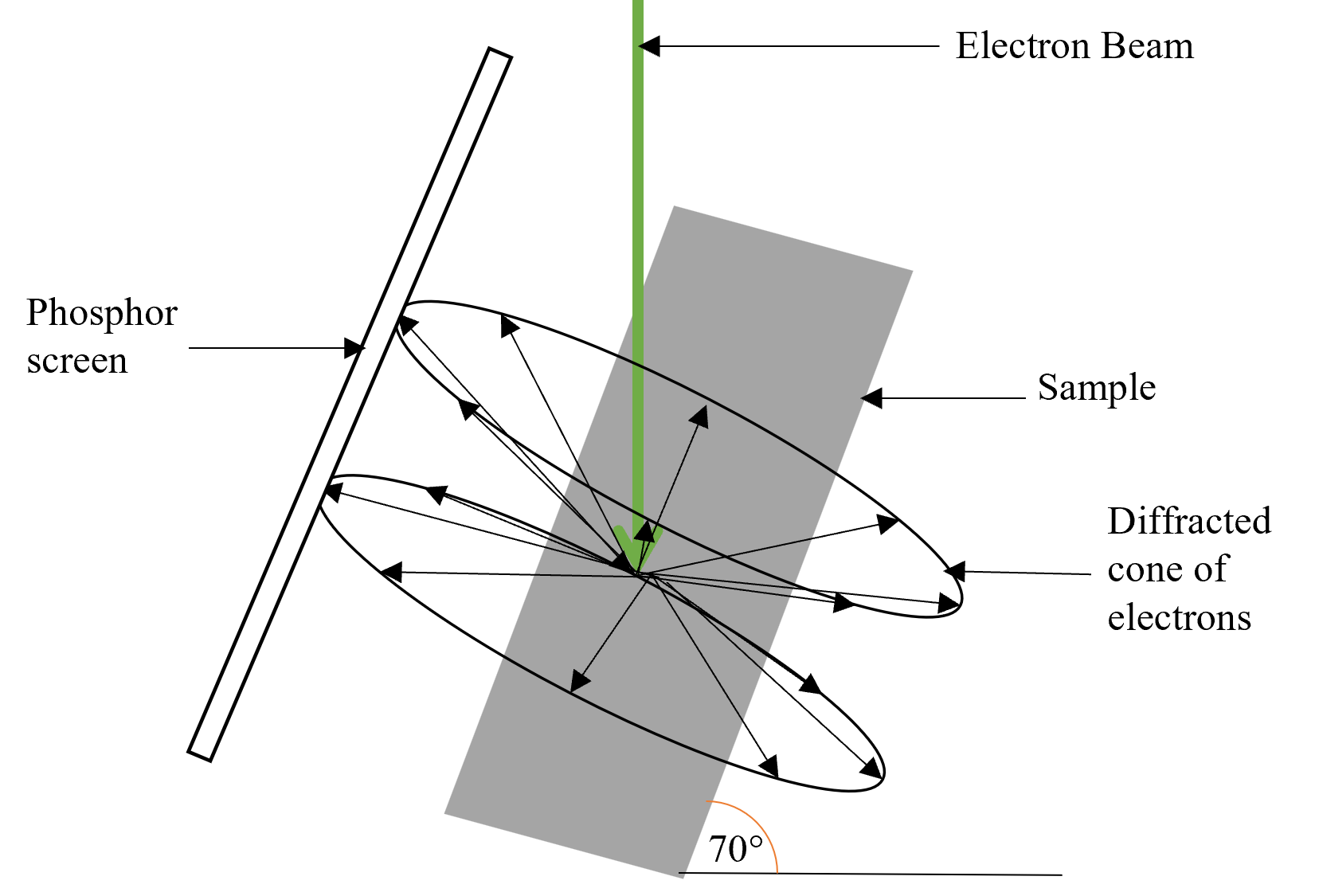}
    \caption{The formation of an EBSD pattern. An EBSD pattern is produced when the beam of electrons strikes a crystal plane in a way that satisfies the Bragg equation [2], producing a pair of large angle cones which, when intersecting a phosphor screen, produces a bright band with dark edges \cite{code1}.}
    \label{fig:EBSPFormation}
\end{figure}

As illustrated in \ref{fig:SetUp}, the sample is scanned by the electron beam, with a 2-dimensional (2D) EBSD pattern being taken at each scanned location, producing a 4-dimensional (4D) data-set. EBSD patterns are characteristic of the phase and orientation of a crystalline material, containing multiple bands from the strongest reflecting planes, as demonstrated by \ref{fig:EBSPFormation} Prior to EBSD acquisition, some knowledge of the sample is required, such as the chemical composition, which allows the correct dictionaries to be used for the indexing step. 
Indexing is performed on each pattern as it is acquired based on the dictionaries provided. Indexing allows the bands in the EBSD patterns to be identified and characterised based on the known reflections of the possible sample structures, allowing the orientations and phases to be identified \cite{code11}. Once indexing has been completed, maps of the sample can be created to visualise the sample structure. In these indexed maps each pixel corresponds to a probe position, not to be confused with the pixel positions within an acquired EBSD pattern.

Since the first observation of a backscattered diffraction pattern in 1928 \cite{code2} acquisition speed and processing time has significantly improved, going from maximum acquisition speeds of approximately 1500 patterns per second in 2015 using a charge coupled device detector to speeds reach 6000 patterns per second with the implementation of the complementary metal oxide superconductor detector \cite{code3}. Automated indexing is commonplace in commercial systems allowing bands in patterns to be easily identified and indexed. Despite these innovations, the speed of EBSD is limited by the acquisition of patterns.

 \begin{figure*}[t]
    \centering
    \begin{minipage}{\textwidth}
    
    \begin{minipage}{0.17\textwidth}
        \centering
        \includegraphics[width=\textwidth]{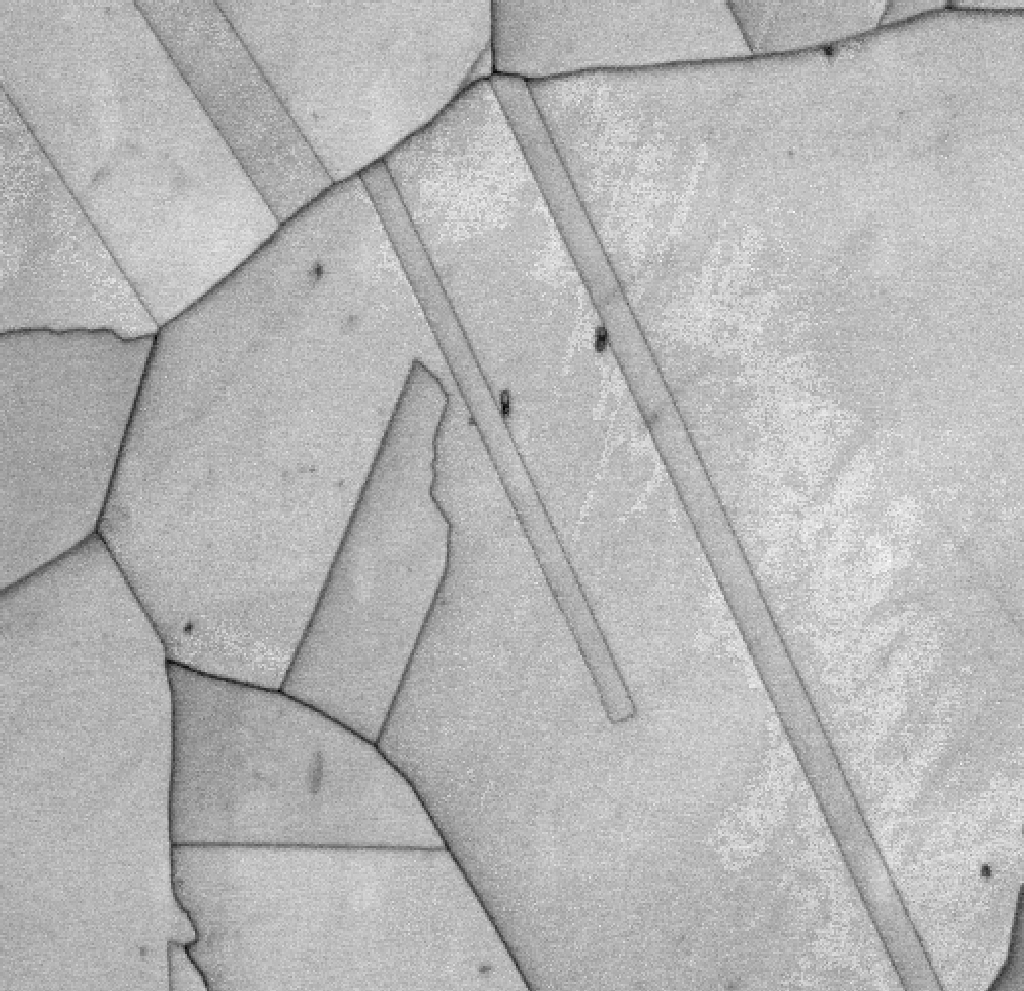}
    \end{minipage}
    \begin{minipage}{0.17\textwidth}
        \centering
        \includegraphics[width=\textwidth]{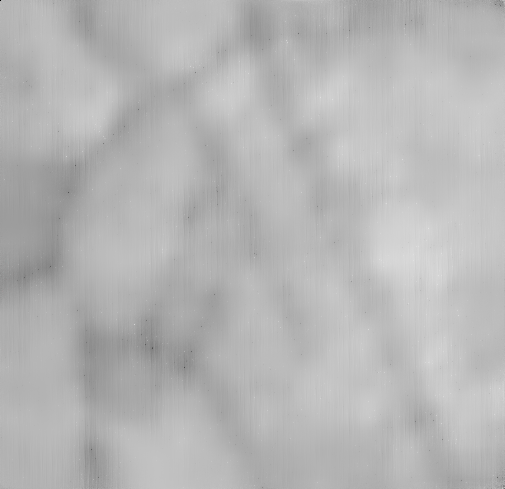}
    \end{minipage}
    \begin{minipage}{0.17\textwidth}
        \centering
        \includegraphics[width=\textwidth]{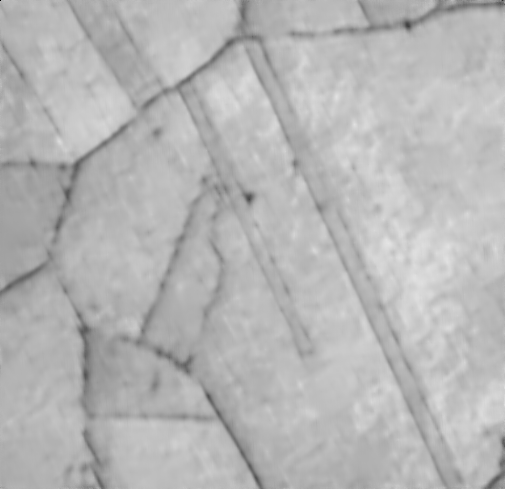}
    \end{minipage}
    \begin{minipage}{0.17\textwidth}
        \centering
        \includegraphics[width=\textwidth]{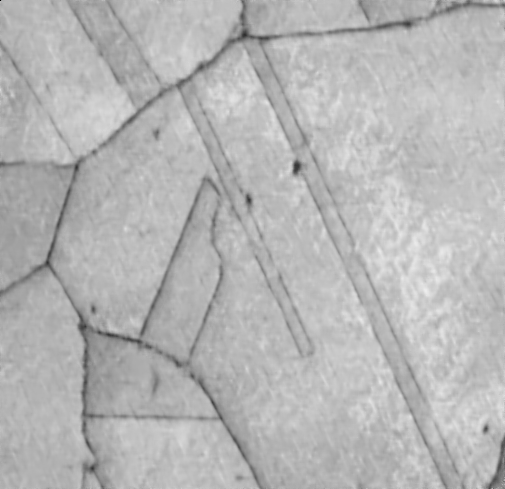}
    \end{minipage}
    \begin{minipage}{0.17\textwidth}
        \centering
        \includegraphics[width=\textwidth]{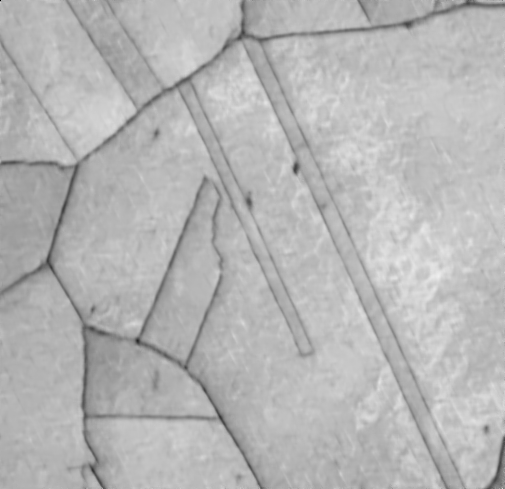}
    \end{minipage}
    \begin{minipage}{0.11\textwidth}
        
    \end{minipage}
    \vspace{1mm}
\end{minipage}
\begin{minipage}{\textwidth}

    \begin{minipage}{0.17\textwidth}
        \centering
        \includegraphics[width=\textwidth]{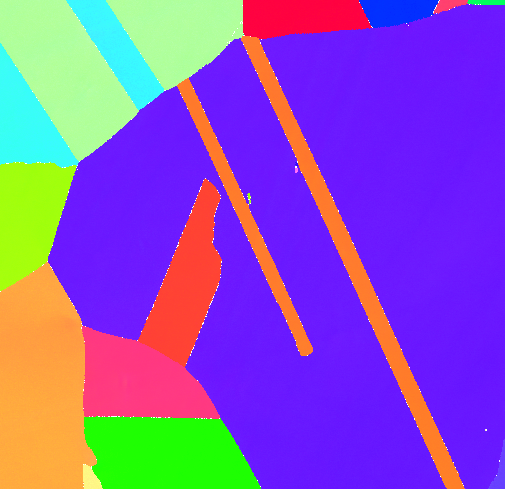}
    \end{minipage}
    \begin{minipage}{0.17\textwidth}
        \centering
        \includegraphics[width=\textwidth]{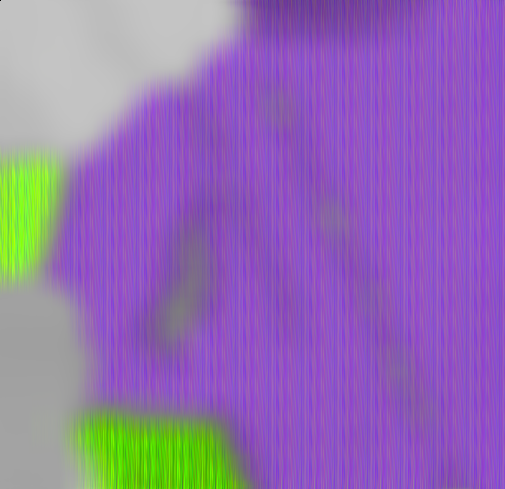}
    \end{minipage}
    \begin{minipage}{0.17\textwidth}
        \centering
        \includegraphics[width=\textwidth]{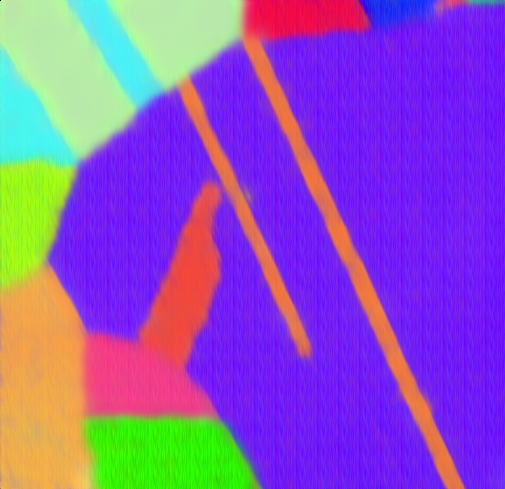}
    \end{minipage}
    \begin{minipage}{0.17\textwidth}
        \centering
        \includegraphics[width=\textwidth]{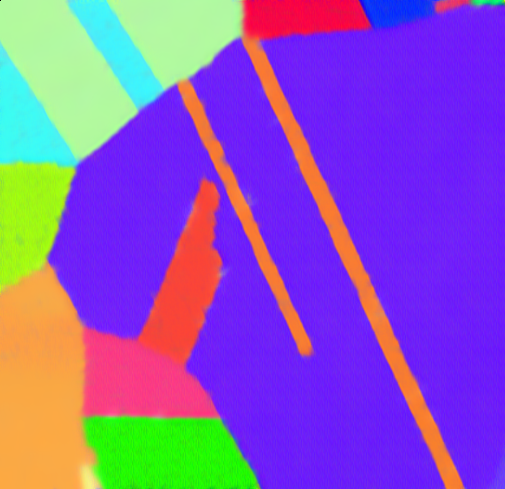}
    \end{minipage} 
    \begin{minipage}{0.17\textwidth}
        \centering
        \includegraphics[width=\textwidth]{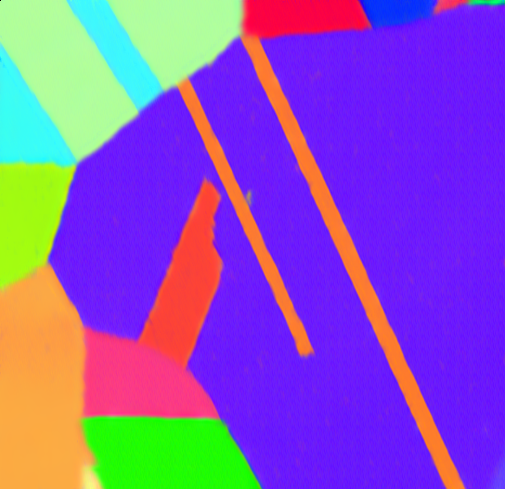}
    \end{minipage} 
    \begin{minipage}{0.12\textwidth}
        \centering
    \includegraphics[width=\textwidth]{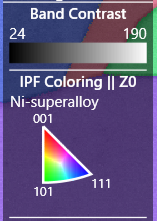}
    \end{minipage}
    \end{minipage}
    \caption{(top) Band contrast map. (bottom) IPF map. From left to right: reference map, reconstructed maps with 1, 5, 10 and 15 \% probe subsampling, a legend describing the band contrast and IPF maps.} \sq \sq
    \label{fig:ReconBCMap}
\end{figure*}

Inspired by compressive sensing theory \cite{code4,code5}, it is believed that EBSD data acquisition speeds could be improved without any significant advances in detector technology being necessary. In recent studies, compressive sensing has been found to be effective in numerous microscopy applications, notably in scanning transmission electron microscopy (STEM) \cite{code6,code7,code8,code9}. Through subsampling signals, data can be acquired at a greater speed and reconstructed using an inpainting algorithm, which could significantly improve the speed of acquisition and processing with regards to EBSD detection. Moreover, in the cases of beam sensitive materials, the effects of beam damage would be greatly reduced. 

In this paper we introduce a new EBSD acquisition approach based on subsampling probe positions; resulting in an incomplete 4D EBSD data. Indexing such data, however, yields band contrast and IPF maps with missing values associated with unsampled probe positions. Our approach is to inpaint the indexed images using Beta Process Factor Analysis (BPFA), \ie a joint blind dictionary learning and sparse coding.
Our inpainting approach is based on a fast GPU implementation of BPFA; allowing significantly faster data acquisition without any significant advances in detector technology being necessary.




\sq \section{Methods}
\sq Assume an EBSD detector with $N_1\times N_2$ pixels recording the measurements from $N_3 \times N_4$ scanning electron probes. Let $ X\in\mathbb{R}^{N_1 \times N_2 \times N_3 \times N_4}$ be the discretised 4-dimensional representation of the EBSD dataset, with $V_k \in \mathbb{R}^{N_3 \times N_4}$ representing each virtual image corresponding to a pair of detector pixel coordinates $k \in \{1,...,N_q\}$ with $N_q = N_1 N_2$ being the total number of pixels in the detector. We assimilate each $V_k$ to its vectorised version $v_k := {\rm vec}(V_k) \in \mathbb{R}^{N_p}$, where $N_p = N_3 N_4$ is the total number of probe positions. We now introduce the subsampling of probe positions to reduce the acquisition time and beam damage. This amounts to subsampling virtual images where the number of acquired spatial locations in the $k^{\rm th}$ virtual image is given as $M < N_p$ acquired over the subsampling set $\Omega \subset \left \{ 1,...,N_{p} \right \}$ with cardinality $\left | \Omega  \right |= M$. This defines our acquisition model as,
\begin{equation}
    y_{k} = P_{\Omega }v_{k}+n_{k} \in \mathbb{R}^{N_{p}},
\end{equation}
for every virtual image where $P_{\Omega }\in \left \{ 0,1 \right \}^{N_{p}\times N_{p}}$ is a mask operator with $\left ( P_{\Omega }v_{k} \right )_{j}= 0$ if $j\in \Omega $ and $\left ( P_{\Omega }v_{k} \right )_{j} = 0$ otherwise. The notable takeaway is that $\Omega$ is the same for all virtual images.
\vspace{0.5mm}

\sq Once the measurements have been acquired, indexing is performed, there is no impact on the indexing step through probe subsampling: missing probe positions are ignored during indexing. The subsampled data is then inpainted using BPFA, \ie a blind dictionary learning approach. For a full explanation of this recovery method refer to Section 2.2 of \cite{code9}.
 

\sq \section{Simulations}
\sq Presented here is one sub-sampling application using a 1024 x 704 pixel EBSD data set of Ni-superalloy. This data was acquired at an acquisition speed of approximately 1600 patterns per second, giving an acquisition time of 7 minutes 24 seconds. 

\begin{figure}
    \centering
    	\begin{tikzpicture}
		\begin{axis}[%
			width=3in,
			height=1.5in,
			at={(0.762in,0.486in)},
			scale only axis,
			xmin=0,
			xmax=25,
			xlabel={Subsampling Ratio (\%)},
			xtick={1,5,10,15,20,25},
			xticklabels={1,5,10,15,20,25},
			xlabel style={at = {(0.5,-0.08)},font=\fontsize{12}{1}\selectfont},
			ticklabel style={font=\fontsize{10}{1}\selectfont},
			ymin=0.4,
			ymax=1,
			ytick={0.4, 0.5, 0.6, 0.7, 0.8, 0.9, 1},
			yticklabels={0.4, 0.5, 0.6, 0.7, 0.8, 0.9, 1},
			ylabel={SSIM},
			ylabel style={at = {(-0.08,0.5)},font=\fontsize{12}{1}\selectfont},
			axis background/.style={fill=white},
			legend style={at={(0.95,0.05)},anchor=south east,legend cell align=left,align=left,fill=none,draw=none,font=\fontsize{6}{1}\selectfont},
			legend entries={{Inverse Pole Figure},{Band Contrast}},
			]
			\addlegendimage{mark=+,black,solid,very thick},
			\addlegendimage{mark=o,blue,solid,very thick},
			\addplot [mark=+,color=black,solid,line width = 1.5, forget plot]
			table[row sep=crcr]{
                1       0.459604581\\
                5       0.676646475\\
                10     0.812721214\\
                15     0.895751802\\
                20     0.920830018\\
                25     0.934234203\\
            };
			\addplot [mark=o,color=blue,solid,line width = 1.5, forget plot]
			table[row sep=crcr]{
                1       0.687488232\\
                5       0.768031718\\
                10     0.776415511\\
                15     0.821453074\\
                20     0.849217138\\
                25     0.858081061\\
			};
		\end{axis}
	\end{tikzpicture}%
    \caption{SSIM of Post-Indexing Reconstructed IPF and Band Contrast Maps; a PSNR plot has also been created for these data points.}
    \label{fig:IPFSSIM}
\end{figure}
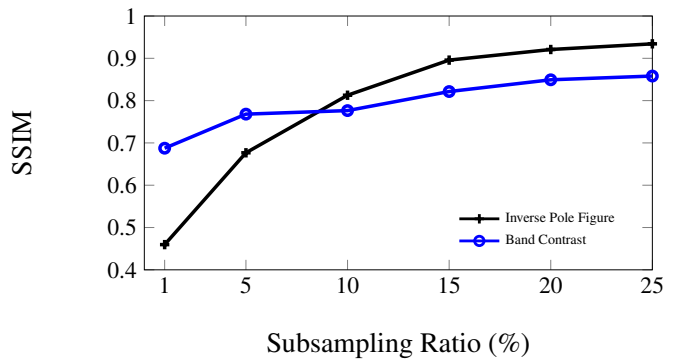

The inpainting algorithm investigated was BPFA \cite{code10}, with band contrast, and IPF maps being investigated. 
 A 512 x 512 pixel subset of the data was subsampled and reconstructed, shown in Figure \ref{fig:ReconBCMap}. The structural similarity (SSIM) results of the reconstructions are shown in \ref{fig:IPFSSIM}. 
 The IPF map had a significantly lower SSIM values at 1 and 5 \% sampling ratios. Meanwhile, the band contrast map had more consistent SSIM values, ranging from 0.69 to 0.86 from sampling ratios of 1 to 25 \%. 
 Acquisition speed of the data is theoretically proportional to the sampling ratio, with a sampling ratio of 10\% taking 10\% of the acquisition time of a full scan. 

\sq \section{Conclusions}
\sq A proof of concept for subsampling scanning probe positions in the context of EBSD data acquisition using scanning electron microscopy have been discussed. Further investigation into this method will allow for the extension of reconstruction into additional map types. 
A second subsampling scanning probe position technique will also be investigated, inpainting the EBSD patterns which can then be indexed alongside the acquired patterns to produce a full 4D dataset. 

\vfill\pagebreak





\bibliographystyle{IEEEtran}
\bibliography{references}
\end{document}